\documentstyle[preprint,aps]{revtex}
\tighten
\begin{document}
\preprint{THEF-NYM 93.05}
\draft
\hyphenation{Nijmegen}
\hyphenation{Rijken}
\hyphenation{Stich-ting Fun-da-men-teel On-der-zoek Ma-te-rie}
\hyphenation{Ne-der-land-se Or-ga-ni-sa-tie We-ten-schap-pe-lijk}

\title{Construction of high-quality \protect\bbox{\!N\!N}
       potential models}

\author{V.G.J.\ Stoks}
\address{School of Physical Sciences, The Flinders University of South 
         Australia, Bedford Park, South Australia 5042, 
         Australia~\cite{perm} \\
         and Institute for Theoretical Physics, University of Nijmegen,
         Nijmegen, The Netherlands}
\author{R.A.M.\ Klomp, C.P.F.\ Terheggen, and J.J.\ de Swart}
\address{Institute for Theoretical Physics~\cite{email}, 
         University of Nijmegen, Nijmegen, The Netherlands}

\date{submitted to Phys.\ Rev.\ C}
\maketitle

\begin{abstract}
We present an updated version (Nijm93) of the Nijmegen soft-core
potential, which gives a much better description of the $np$ data
than the older version (Nijm78). The $\chi^{2}$ per datum is 1.87.
The configuration-space and momentum-space versions of this potential
are exactly equivalent; a unique feature among meson-theoretical
potentials.
We also present three new $N\!N$ potential models: a non-local 
Reid-like Nijmegen potential (Nijm~I), a local version (Nijm~II), 
and an updated regularized version (Reid93) of the Reid soft-core 
potential. These three potentials all have a nearly optimal $\chi^{2}$ 
per datum and can therefore be considered as alternative partial-wave 
analyses.
All potentials contain the proper charge-dependent one-pion-exchange 
tail.
\end{abstract}

\widetext

\section{INTRODUCTION}
\label{intro}
In the past many nucleon-nucleon ($N\!N$) potentials were constructed,
which were supposed to fit the $N\!N$ scattering data available at 
the time of construction. 
The older models, from the 1950s and 1960s, are no longer 
suitable for describing the present set of more numerous and much 
more accurate data without refitting the parameters. 
Out of the various potential models constructed in the 1970s, the 
better ones fitted the data with $\chi^{2}/N_{\rm data}$ of about 2, 
where $N_{\rm data}$ denotes the number of $N\!N$ scattering data
available at that time in the 0--350 MeV energy range. 
The potentials constructed in the 1980s have only slightly improved 
on this in the sense that, although they have been fitted to try to 
describe the newer and much more accurate data, these models still have 
$\chi^{2}/N_{\rm data}\approx2$. This number should be compared with 
$\chi^{2}_{\rm min}/N_{\rm data}=0.99$, obtained in the recently 
finished Nijmegen $N\!N$ multienergy partial-wave analysis~\cite{St93b} 
(PWA93) of all $pp$ and $np$ scattering data below 350 MeV.
On statistical grounds, $\chi^{2}_{\rm min}/N_{\rm data}\approx1$ is 
about the best one can expect to get in partial-wave analyses or 
for potential models.

In a recent paper~\cite{St93a}, we investigated the quality with 
respect to the $pp$ scattering data below 350 MeV of a number of $N\!N$ 
potentials that had appeared in the literature. We found that only a 
few of the potential models we investigated are of a satisfactory 
quality. These models are the Reid soft-core potential~\cite{Rei68} 
Reid68, the Nijmegen soft-core potential~\cite{Nag78} Nijm78, and the 
new Bonn $pp$ potential~\cite{Hai89} Bonn89. The latter is a 
readjustment of the momentum-space full Bonn potential~\cite{Mac87},
in order to fit the $pp$ data.
If we do not consider the very low-energy (0--2 MeV) $pp$ data, also 
the parametrized Paris potential~\cite{Lac80} Paris80 gives a 
satisfactory description of the data. 
The results of Ref.~\cite{St93a} indicate that, at present, the best 
potential models have $\chi^{2}/N_{pp}\gtrsim1.9$, where $N_{pp}$ 
denotes the number of $pp$ scattering data. Moreover, only models which
have explicitly included the $pp$ data in their fit belong to this
category. Potential models which have been fitted only to the $np$ data
often give a poor description of the $pp$ data, even after
applying the necessary corrections for the Coulomb interaction.
In Ref.~\cite{St93a} we have demonstrated that most $np$ potentials
unfortunately do not automatically fit the $pp$ data, a fact which has
been generally overlooked. Any $N\!N$ potential should be fitted to
the $pp$ data as well as to the $np$ data in order to be able to 
describe all $N\!N$ scattering data.

Over the last decade the quality of the $np$ data has increased
considerably. Consequently, the older potentials (Reid68, Nijm78,
Paris80) do not fit these data very well. Also, the much newer Bonn 
potentials already needed revisions and updates~\cite{Hai89,Mac89}.
In this paper we present updates of the Nijm78 and Reid68 potentials, 
denoted by Nijm93 and Reid93, respectively. 
Because our analysis of the $np$ data (and hence our careful scrutiny 
of the $np$ data) has only recently been finished~\cite{St93b}, we 
originally constructed an update (Nijm92$pp$) of the Nijm78 potential 
for the $pp$ data only. 
This $pp$ potential was used in our earlier preliminary $np$ 
analyses~\cite{Klo91,Klo92,St93c} to parametrize the isovector partial 
waves. It has $\chi^{2}/N_{pp}=1.4$, which is not as good as the 
Nijmegen PWA93. 
One can wonder whether it is at all possible to construct a new class 
of potential models which fit the $N\!N$ data with the almost perfect 
$\chi^{2}/N_{\rm data}\approx1$. The answer turns out to be affirmative. 
This could already be surmised from the Nijmegen PWA93, because this 
analysis is in essence an energy-dependent potential fitted to the 
scattering data. (The reason for us using an energy-dependent potential 
is nothing more than just convenience.) 
In the partial-wave analysis we need 39 parameters to reach 
$\chi^{2}/N_{\rm data}=0.99$, whereas a conventional potential model 
typically has only 10--15 free parameters. It is therefore perhaps not 
surprising that the 15-parameter update (Nijm93) of the Nijm78 potential, 
which fits the $N\!N$ data with $\chi^{2}/N_{\rm data}=1.87$, cannot 
compete in quality with the Nijmegen PWA93. To obtain a high-quality
potential we decided some years ago to follow a different approach.

Because the Nijm92$pp$ potential already gives a reasonable description 
of the $pp$ data, this model forms the basis for the construction of a 
high-quality potential, which {\it can\/} compete with the Nijmegen PWA93. 
We adjust in each partial wave separately only a few of the parameters 
of this potential~\cite{remark}.
This way we will be able to construct a potential model which fits 
the data with $\chi^{2}/N_{\rm data}\approx1$. 
The resulting Reid-like potential Nijm~I gives a very good fit to the 
data with $\chi^{2}/N_{\rm data}=1.03$. 

The Nijm~I potential contains momentum-dependent terms (as do the Nijm78 
and Nijm93 potentials), which in configuration space give rise to a 
non-local structure $(\Delta\varphi(r)+\varphi(r)\Delta)$ to the potential.
We also constructed a purely local Nijm~II potential, where these 
momentum-dependent terms were intentionally omitted. This local
potential Nijm~II gives an equally good fit to the data as the 
non-local potential Nijm~I. Finally, we constructed 
a regularized update of the Reid68 potential~\cite{Rei68}, called
Reid93. This Reid93 model is also a local potential and fits the
scattering data very well.
These latter three potential models are in a sense also alternative 
partial-wave analyses, because they have roughly the same number of 
fit parameters as our Nijmegen PWA93, these parameters were fitted 
to the same database, and the potential models achieve nearly the same 
values of $\chi^{2}_{\rm min}$ as the Nijmegen PWA93 (i.e., close to the 
expectation value). Hence, the differences between, e.g., the phase
parameters of these models provide an indication for the systematic
error in the Nijmegen partial-wave analyses.

In Sec.~\ref{outline} we briefly discuss some general features of 
$N\!N$ potentials. In Sec.~\ref{structure} we give more details 
regarding the explicit form of the potentials used in this work. 
Two of these potentials are based on the original Nijm78 potential, 
whereas the third is a regularized update of the Reid68 potential in 
the sense that also in this new model each partial wave is parametrized 
by a number of Yukawa functions. 
In Sec.~\ref{results} we discuss the fitting procedure and the
potentials are presented in more detail.

\section{GENERAL OUTLINE}
\label{outline}
The $N\!N$ potential can be described in momentum space and in 
configuration space. Since it is difficult to solve the full
four-dimensional scattering equation, it has become common practice
first to make a reduction to a three-dimensional scattering equation.
Various choices are possible, and it is important to note that the
potential derived within the chosen reduction scheme should 
{\it only\/} be used in the scattering equation corresponding to that 
particular reduction scheme. These three-dimensional scattering 
equations can always be written in the form of the momentum-space 
version of the Lippmann-Schwinger equation. If the kinematics is treated
relativistically, this is called the relativistic Lippmann-Schwinger
equation. In configuration space, the differential form of this
integral equation is the Schr\"odinger equation.

The configuration-space potentials are to be used either in the
nonrelativistic or the relativistic Schr\"odinger equation
\begin{equation}
    (\Delta+k^{2})\psi = 2M_{r}V\,\psi \ ,  \label{Schroed}
\end{equation}
where $\Delta$ is the Laplacian, and where (non)relativistic refers 
to the kinematics. For nonrelativistic kinematics the relation 
between the center-of-mass energy $E$ and the center-of-mass momentum 
squared $k^{2}$ reads $E=k^{2}/2M_{r}$, whereas for relativistic 
kinematics it reads 
$E=\sqrt{k^{2}+M^{2}_{1}}+\sqrt{k^{2}+M^{2}_{2}}-M_{1}-M_{2}$. 

The earliest potential models were configuration-space potentials to
be used in the nonrelativistic Schr\"odinger equation. They were
phenomenological or semiphenomenological parameterizations, based on
a general form for the potential. The potential must be invariant under 
rotations, reflections, and time reversal, and can be written~\cite{Oku58}
as the sum of 6 independent terms, $V=\sum_{i=1}^{6}V_{i}P_{i}$.
A common choice for the 6 operators $P_{i}$ in configuration space is
\begin{equation}
  \begin{array}{l}
   P_{1}=1,                                           \\[0.2cm]
   P_{2}=\mbox{\boldmath $\sigma$}_{1}\!\cdot\!
         \mbox{\boldmath $\sigma$}_{2},               \\[0.2cm]
   P_{3}=S_{12}=3(\mbox{\boldmath $\sigma$}_{1}\!\cdot\!\hat{\bf r})
                 (\mbox{\boldmath $\sigma$}_{2}\!\cdot\!\hat{\bf r})
                -(\mbox{\boldmath $\sigma$}_{1}\!\cdot\!
                  \mbox{\boldmath $\sigma$}_{2}),     \\[0.2cm]
   P_{4}={\bf L\cdot S},                              \\[0.2cm]
   P_{5}=Q_{12}={\textstyle\frac{1}{2}}
         [(\mbox{\boldmath $\sigma$}_{1}\!\cdot\!{\bf L})
          (\mbox{\boldmath $\sigma$}_{2}\!\cdot\!{\bf L})+
          (\mbox{\boldmath $\sigma$}_{2}\!\cdot\!{\bf L})
          (\mbox{\boldmath $\sigma$}_{1}\!\cdot\!{\bf L})], \\[0.2cm]
   P_{6}={\textstyle\frac{1}{2}}
        (\mbox{\boldmath $\sigma$}_{1}-\mbox{\boldmath $\sigma$}_{2})
        \cdot{\bf L} .
   \end{array}     \label{Pconfig}
\end{equation}
These operators are also frequently referred to as the central, 
spin-spin, tensor, spin-orbit, quadratic spin-orbit, and antisymmetric 
spin-orbit operators, respectively.
For identical-particle scattering, the antisymmetric spin-orbit 
operator $P_{6}$ cannot contribute, whereas $V_{6}$ vanishes when
charge independence is assumed (which is usually the case for $N\!N$
potential models).
In general~\cite{Oku58}, each potential form $V_{i}$ in configuration
space is a function of $r^{2}$, and of the operators $p^{2}$ and $L^{2}$. 
In most approaches one only keeps the dependence on $r^{2}$, while the 
$p^{2}$ dependence (when included) is often only present in a linear
way in the central potential $V_{1}$. 
The inclusion of the $Q_{12}$ operator was found to be necessary,
because otherwise it was impossible to describe simultaneously the
$^{1}S_{0}$ and $^{1}D_{2}$ phase shifts using the same static potential.
The presence of the operator $Q_{12}$ in the potential can to a certain
extent be simulated by introducing non-local potentials~\cite{Bry69}.

In the expansion $\sum_{i=1}^{6}V_{i}P_{i}$, the potential forms $V_{i}$
are generally assumed to be the same in all partial waves. The potential 
differences between the partial waves are dictated by the differences in 
the expectation values of the operators $P_{i}$ in these partial waves.
The Reid68 potential~\cite{Rei68}, however, is based on a quite different 
approach. Rather than having 6 potential forms $V_{i}$ which are the same
for all partial waves, now each partial wave is parametrized separately.
The potential forms $V_{i}$ therefore not only depend on $r^{2}$ and 
$L^{2}$, but also on $S^{2}$ and $J^{2}$.
In this paper we present some potentials based on this approach that
each partial wave is parametrized independently.
We refer to these models as Reid-like models.

With the discovery of the heavy mesons in the 1960s, it became common
practice to write the potential as a sum over one-boson-exchange (OBE)
potentials. The expressions for these OBE potentials are usually
derived in momentum space. Introducing momentum vectors
\[
   {\bf k}={\bf p}_{f}-{\bf p}_{i}, \ \
   {\bf q}={\textstyle\frac{1}{2}}({\bf p}_{f}+{\bf p}_{i}), \ \
   {\bf n}={\bf q}\times{\bf k}\ ,
\]
in terms of initial (${\bf p}_{i}$) and final (${\bf p}_{f}$)
momenta, the equivalent in momentum space to Eq.~(\ref{Pconfig}) reads
\begin{equation}
  \begin{array}{l}
   P_{1}=1                                            \\[0.2cm]
   P_{2}=\mbox{\boldmath $\sigma$}_{1}\!\cdot\!
         \mbox{\boldmath $\sigma$}_{2},               \\[0.2cm]
   P_{3}=(\mbox{\boldmath $\sigma$}_{1}\!\cdot\!{\bf k})
         (\mbox{\boldmath $\sigma$}_{2}\!\cdot\!{\bf k})
        -{\textstyle\frac{1}{3}}{\bf k}^{2}
         (\mbox{\boldmath $\sigma$}_{1}\!\cdot\!
          \mbox{\boldmath $\sigma$}_{2}),             \\[0.2cm]
   P_{4}={\textstyle\frac{i}{2}}
        (\mbox{\boldmath $\sigma$}_{1}+\mbox{\boldmath $\sigma$}_{2})
        \cdot{\bf n},                                 \\[0.2cm]
   P_{5}=(\mbox{\boldmath $\sigma$}_{1}\!\cdot\!{\bf n})
         (\mbox{\boldmath $\sigma$}_{2}\!\cdot\!{\bf n}), \\[0.2cm]
   P_{6}={\textstyle\frac{i}{2}}
        (\mbox{\boldmath $\sigma$}_{1}-\mbox{\boldmath $\sigma$}_{2})
        \cdot{\bf n} .
   \end{array}     \label{Pmom}
\end{equation}
The potential forms $V_{i}$ in momentum space are functions of 
${\bf k}$, ${\bf q}$, ${\bf n}$, and the energy.
Although Eq.~(\ref{Pmom}) provides an adequate set of 6 linearly 
independent operators, the $Q_{12}$ operator in configuration space is 
{\it not\/} the exact Fourier transform of the
$(\mbox{\boldmath $\sigma$}_{1}\!\cdot\!{\bf n})
(\mbox{\boldmath $\sigma$}_{2}\!\cdot\!{\bf n})$ operator in momentum 
space. This is of importance if we want both the momentum-space and 
the configuration-space versions to produce exactly the same phase shifts 
and bound states, which is only possible when the configuration-space
version is the exact Fourier transform of the momentum-space version,
and vice versa. This implies~\cite{Rij91} that we have to use the 
inverse Fourier transform of the $Q_{12}$ operator; i.e., the potential 
contribution $(\mbox{\boldmath $\sigma$}_{1}\!\cdot\!{\bf n})
(\mbox{\boldmath $\sigma$}_{2}\!\cdot\!{\bf n})\,V_{5}({\bf k}^{2})$ 
is to be replaced by
\begin{equation}
  P_{5}V_{5}({\bf k}^{2}) - P'_{5}\int_{\infty}^{{\bf k}^{2}}
        \!d{\bf k}'^{2}V_{5}({\bf k}'^{2})   \ , \label{Q12replace}
\end{equation}
where
\begin{eqnarray}
    P'_{5} &=& [(\mbox{\boldmath $\sigma$}_{1}\!\cdot\!{\bf q})
                (\mbox{\boldmath $\sigma$}_{2}\!\cdot\!{\bf q})
            -{\bf q}^{2}(\mbox{\boldmath $\sigma$}_{1}\!\cdot\!
                         \mbox{\boldmath $\sigma$}_{2})]  \nonumber\\
           & & -{\textstyle\frac{1}{4}}
               [(\mbox{\boldmath $\sigma$}_{1}\!\cdot\!{\bf k})
                (\mbox{\boldmath $\sigma$}_{2}\!\cdot\!{\bf k})
               -{\bf k}^{2}(\mbox{\boldmath $\sigma$}_{1}\!\cdot\!
                            \mbox{\boldmath $\sigma$}_{2})]  \ .
\end{eqnarray}
Other restrictions imposed on the momentum-space potential forms $V_{i}$ 
in that case are that they should not depend on the energy, while the 
${\bf q}$ dependence should be of second order at most (see also below).

When the potentials are evaluated in momentum space and then Fourier
transformed to configuration space, they are usually first regularized
to remove the singularities at the origin. This can be achieved by
introducing a form factor $F({\bf k}^{2})$.
A typical Fourier transform, encountered in transforming the
momentum-space potential to configuration space, then reads
\begin{eqnarray}
  \int\frac{d^{3}k}{(2\pi)^{3}}\,\frac{e^{i{\bf k}\cdot{\bf r}}}
     {{\bf k}^{2}+m^{2}} \left({\bf k}^{2}\right)^{n}F({\bf k}^{2})
  &\equiv&\frac{m}{4\pi}(-m^{2})^{n}\phi^{n}_{C}(r)       \nonumber\\
  &=&\frac{m}{4\pi}(-\mbox{\boldmath $\nabla$}^{2})^{n}\phi^{0}_{C}(r)\ .
             \label{phiC}
\end{eqnarray}
The results for various frequently used choices are:
\newline
(i) No form factor at all, $F({\bf k}^{2})=1$. This yields the 
familiar Yukawa function
\begin{equation}
     \phi^{0}_{C}(r)=e^{-mr}/mr   \ ,            \label{Yukawa}
\end{equation}
and the singularities at the origin are still present;  
\newline
(ii) Monopole form factor, $F({\bf k}^{2})=(\Lambda^{2}-m^{2})/
(\Lambda^{2}+{\bf k}^{2})$, normalized such that at the pole 
$F(-m^{2})=1$. This yields
\begin{equation}
     \phi^{0}_{C}(r)=\left[e^{-mr}-e^{-\Lambda r}\right]/mr   \ ;
\end{equation}
(iii) Dipole form factor, $F({\bf k}^{2})=(\Lambda^{2}-m^{2})^{2}/
(\Lambda^{2}+{\bf k}^{2})^{2}$, yielding
\begin{equation}
     \phi^{0}_{C}(r)=\left[e^{-mr}-e^{-\Lambda r} 
     \left(1+\frac{\Lambda^{2}-m^{2}}{2\Lambda^{2}}\Lambda r\right)  
              \right] \mbox{\LARGE /}mr       \ ; \label{phidipole}
\end{equation}
(iv) Exponential form factor, $F({\bf k}^{2})=
e^{-{\bf k}^{2}/\Lambda^{2}}$, yielding
\begin{eqnarray}
     \phi^{0}_{C}(r)=e^{m^{2}/\Lambda^{2}}
     \biggl[&&e^{-mr}{\rm erfc}
          \left(\frac{m}{\Lambda}-\frac{\Lambda r}{2}\right) \nonumber\\
     &&    - e^{mr}{\rm erfc}
          \left(\frac{m}{\Lambda}+\frac{\Lambda r}{2}\right)\biggr]
            \mbox{\LARGE /}2mr \ ,
\end{eqnarray}
where erfc$(x)$ is the complementary error function
\[  {\rm erfc}(x)=\frac{2}{\sqrt{\pi}}\int_{x}^{\infty}\!
      dt\, e^{-t^{2}}   \ .  \]
We follow the normalization of Ref.~\cite{Nag78}. This means that for 
the exponential form factor $F(0)=1$. 

Because ${\bf k}^{2}$ can be written as $({\bf k}^{2}+m^{2})-m^{2}$,
we find that in the absence of a form factor
\begin{equation}
  \phi^{1}_{C}(r)=\phi^{0}_{C}(r)-4\pi\delta^{3}(m{\bf r}) \ , 
                                                      \label{phi1C}
\end{equation}
When there is a form factor, this relation still holds, but the 
$\delta$-function contribution is smeared out.

Using our definition (\ref{phiC}), the Fourier transforms for the 
tensor and spin-orbit potentials can be simply expressed in terms of 
derivatives of the central function, i.e.,
\begin{eqnarray}
  \phi^{0}_{T}(r) &=& \frac{1}{3m^{2}}\,r\frac{d}{dr}\left(\frac{1}{r}
            \frac{d}{dr}\right)\, \phi^{0}_{C}(r) \ , \nonumber\\
  \phi^{0}_{SO}(r)&=&-\frac{1}{m^{2}} \frac{1}{r}\frac{d}{dr}\, 
                                    \phi^{0}_{C}(r) \ . \label{phiTSO}
\end{eqnarray}
In order to ensure regularity at the origin for the tensor and 
spin-orbit functions, one must choose at least the dipole or exponential 
form factor. In that case, the tensor function also vanishes at the 
origin, as it should.
 
The presence of explicit momentum-dependent terms in the momentum-space
potential gives rise to non-local structures in the potential in 
configuration space.
The ${\bf q}^{2}$ terms pose no difficulties for the configuration-space
potential as long as they are linear in ${\bf q}^{2}$.
The typical Fourier transform of such a term is given by
\begin{eqnarray}
  \int\frac{d^{3}k}{(2\pi)^{3}}\,\frac{e^{i{\bf k}\cdot{\bf r}}}
     {{\bf k}^{2}+m^{2}} && \left({\bf q}^{2}+{\textstyle\frac{1}{4}}
                   {\bf k}^{2}\right) F({\bf k}^{2}) \nonumber\\
   =&& -\frac{m}{4\pi} \left[\Delta\frac{\varphi(r)}{2M_{r}}+
            \frac{\varphi(r)}{2M_{r}}\Delta\right] \ , \label{nonlocal}
\end{eqnarray}
where $\varphi(r)=M_{r}\phi^{0}_{C}(r)$. It is well known how to 
handle such a $(\Delta\varphi+\varphi\Delta)$ term~\cite{Gre62}.
The absence of ${\bf q}^{2}$ terms in the momentum-space potential will
result in a radially local configuration-space potential. 

The three new potential models (Nijm93, Nijm~I, and Nijm~II) presented 
in this paper are based on the original Nijm78 potential with the 
exponential form factor, whereas the update of the Reid68 potential 
(Reid93) is regularized using a dipole form factor.

\section{STRUCTURE OF THE POTENTIALS}
\label{structure}
\subsection{One-pion-exchange potential}
\label{subsec:OPE}
An important feature of the potential models presented in this
paper is that in the one-pion-exchange (OPE) part of the potential, 
we explicitly distinguish between neutral-pion and charged-pion 
exchange. The pion masses are~\cite{PDG90} $m_{\pi^{0}}=134.9739$ 
MeV and $m_{\pi^{\pm}}=139.5675$ MeV. Almost all other potentials
that have appeared in the literature use a mean pion mass. In these
other models the isovector $np$ phase parameters are {\it larger\/} in
magnitude than the corresponding $pp$ phase parameters. By explicitly
including the pion-mass differences exactly the opposite occurs:
the isovector $np$ phase parameters are {\it smaller\/} than the
corresponding $pp$ phase parameters. This is a unique feature of
the potentials presented here.

Defining
\begin{equation}
   V(m)\! = \!\left(\frac{m}{m_{\pi^{\pm}}}\right)^{2} \! m \!
    \left[\phi^{0}_{T}(m,r)S_{12}+{\textstyle\frac{1}{3}}
          \phi^{1}_{C}(m,r)(\mbox{\boldmath $\sigma$}_{1}\!
          \cdot\!\mbox{\boldmath $\sigma$}_{2})\right]  , \label{VOPE}
\end{equation}
the OPE potential for $pp$ scattering is given by
\begin{equation}
      V_{\rm OPE}(pp)=f_{\pi}^{2} V(m_{\pi^{0}})    \ , \label{Vpp}
\end{equation}
whereas for $np$ scattering it reads
\begin{equation}
      V_{\rm OPE}(np)=-f_{\pi}^{2} V(m_{\pi^{0}})
                  \pm 2f_{\pi}^{2} V(m_{\pi^{\pm}}) \ , \label{Vnp}
\end{equation}
where the plus (minus) sign corresponds to total isospin $I=1\, (0)$.
The scaling mass $m_{\pi^{\pm}}$ in $V(m)$ is introduced in order to 
make the pseudovector coupling constant $f_{\pi}$ dimensionless. 
It is conventionally chosen to be equal to the charged-pion mass.
The explicit distinction between neutral and charged pions implies that 
the isovector $np$ and $pp$ OPE potentials are different, and so charge 
independence is broken. In our present models we assume, however, that 
the pion-nucleon coupling constants obey charge independence.

\subsection{Nijmegen potential}
\label{subsec:Nijm}
In this section we briefly discuss the structure of the Nijmegen
potential. More details can be found in Refs.~\cite{Nag78,Mae89}. 
The basic functions are the one-boson-exchange (OBE) potential 
functions with momentum-dependent central terms and exponential 
form factors. The meson exchanges we include are those due to 
pseudoscalar mesons ($\pi, \eta, \eta'$), vector mesons 
($\rho, \omega, \phi$), and scalar mesons ($a_{0}, f_{0}, \epsilon$). 
Here we use the modern nomenclature for the scalar mesons, i.e., 
$a_{0}(983)$ corresponds to the $\delta$ of Ref.~\cite{Nag78}, and 
$f_{0}(975)$ to the $S^{\star}$. The $\epsilon$ meson would correspond 
to an $f_{0}(760)$. No such meson is listed by the Particle Data 
Group~\cite{PDG90}; however, a recent analysis of the 
$\pi N\rightarrow\pi^{+}\pi^{-}N$ reaction~\cite{Sve92} provides
evidence for a scalar--isoscalar resonant state $0^{++}(750)$. 
In the Nijmegen potentials the $\epsilon$ meson corresponds to a 
broad meson (see below) where the pole in its propagator is chosen 
to correspond to the pole position in the complex energy plane of 
the isoscalar $\pi\pi$ $S$ wave~\cite{Pro73}. 
Here we will retain the name of $\epsilon$ meson. 
The aforementioned meson exchanges can be identified with the 
dominant parts of the lowest-lying meson trajectories in the complex 
$J$ plane. We furthermore include the dominant $J=0$ parts of the 
Pomeron, and of the $f_{2}, f'_{2}$, and $a_{2}$ tensor-meson
trajectories. They give rise to Gaussian potentials. 

The meson propagators including the exponential form factor read
\begin{equation}
   \Delta({\bf k}^{2},m^{2},\Lambda^{2})=\frac{1}{{\bf k}^{2}+m^{2}}
                 \ e^{-{\bf k}^{2}/\Lambda^{2}}     \ .
\end{equation}
For the Pomeron-type exchanges we have
\begin{equation}
      \Delta({\bf k}^{2},m^{2}_{p})=\frac{1}{M_{p}^{2}}
                 \ e^{-{\bf k}^{2}/4m^{2}_{p}}     \ ,
\end{equation}
where $m_{p}$ has the dimension of a mass and will be called 
the Pomeron mass, and $M_{p}$ is a scaling mass, chosen to be 
the proton mass. 
The different potential forms are evaluated in momentum space and the 
resulting expressions are essentially those of Refs.~\cite{Nag78}
(save some misprints~\cite{misprint}) with the following differences:
(i) We explicitly account for the proton and neutron mass difference;
(ii) the differences between the neutral and charged pion (see 
Sec.~\ref{subsec:OPE}), and between the neutral and charged $\rho$
meson are explicitly included; (iii) we have adjusted the quadratic
spin-orbit operator of the potential in momentum space to include the 
$P'_{5}$ contribution as in Eq.~(\ref{Q12replace}).
The effect of the first modification is obviously rather small.
The second modification (as well as the first) implies that charge
independence is broken in the non-OPE part of the potential as well.
For $pp$ scattering the potential consists of only neutral-meson exchange,
$V_{pp}=V({\rm neutral})$, whereas for $np$ scattering it consists of 
neutral-meson and charged-meson exchange, depending on the total isospin 
as in Eq.~(\ref{Vnp}), so $V_{np}=-V({\rm neutral})\pm2V({\rm charged})$.
This distinction replaces the factor $(\mbox{\boldmath $\tau$}_{1}
\!\cdot\!\mbox{\boldmath $\tau$}_{2})$ used in the old Nijm78 potential.
Finally, the third modification means that we have constructed a
potential which is exactly equivalent in both momentum space and 
configuration space, a unique feature of these Nijmegen potentials.
For example, for the parametrized Paris potential~\cite{Lac80} this
is not true since it uses the same parameters in combination with the 
$Q_{12}$ operator in configuration space as it does with the 
$(\mbox{\boldmath $\sigma$}_{1}\!\cdot\!{\bf n})
 (\mbox{\boldmath $\sigma$}_{2}\!\cdot\!{\bf n})$
operator in momentum space.

Next we briefly discuss the coupling constants. For definitions and
references we again refer to Refs.~\cite{Nag78,Mae89}. The coupling 
constants of the pseudoscalar mesons are related via SU(3) and 
singlet-octet mixing. The octet coupling $f_{\eta_{8}}$ is calculated 
using $\alpha_{P}=0.355$ for the $F/(F+D)$ ratio. For the singlet-octet 
mixing angle we use $\theta_{P}=-23^{\circ}$ to define the physical
coupling constants $f_{\eta}$ and $f_{\eta'}$. This leaves the singlet 
coupling $f_{\eta_{1}}$ and the pion coupling $f_{\pi}$ as free input
parameters. However, in our partial-wave analysis of the $pp$ scattering 
data~\cite{Ber90}, we found for the $pp\pi^{0}$ coupling constant 
$f_{p}^{2}=0.0749(7)$. This value was later confirmed in a combined
partial-wave analysis of all $pp$ and $np$ scattering data, assuming
charge independence for the pion-nucleon coupling 
constants~\cite{Klo91,St93c}. There the value $f_{\pi}^{2}=0.075$ is 
recommended for the pion-nucleon coupling constant at the pion pole. 
This is the value we adopt in our construction of the new Nijmegen 
potentials, and so it is not included as a free parameter.
 
For the vector mesons we assume that the $\rho$ meson is universally 
coupled to the isospin current ($\alpha^{e}_{V}=1$) to define the octet
coupling $g_{V_{8}}$. For the singlet-octet mixing angle we take 
$\theta_{V}=37.5^{\circ}$, which fixes the physical coupling constants 
$g_{\omega}$ and $g_{\phi}$ in terms of $g_{\rho}$ and $g_{V_{1}}$. 
The $\phi$ meson is assumed to have $f_{\phi}\equiv0$. 
The free parameters are now $g_{\rho}$, $g_{V_{1}}$, $f_{\rho}$, and
$f_{\omega}$. 
 
For the scalar mesons we do not apply any constraints for the coupling 
constants, since the singlet-octet mixing angle for the scalar mesons 
is still an unsettled problem (see also Ref.~\cite{Mae89}). The free 
parameters are the $a_{0}$, $f_{0}$, and $\epsilon$ coupling constants. 
 
For simplicity we take a single mass parameter $m_{p}$ for the Pomeron, 
and for the $J=0$ parts of the $f_{2}$, $f'_{2}$, and $a_{2}$ 
tensor-meson trajectories. We use two coupling constants: $g_{a_{2}}$ 
for the isovector $a_{2}$ meson and $g_{p}$ for the isoscalar and
Pomeron exchanges.

For each type of exchange we use an independent cutoff mass, so we have 
three cutoff parameters $\Lambda_{P}$, $\Lambda_{V}$, and $\Lambda_{S}$.
This brings us to a total of 14 free parameters.
  
We conclude this section with a discussion of the treatment of the
broad $\rho$ and $\epsilon$ mesons. The width of a broad meson can be
accounted for~\cite{Sch71,Bin71,Swa78} by replacing the propagator 
$\Delta({\bf k}^{2})=1/({\bf k}^{2}+m^{2})$ of a stable meson by a 
dispersion integral
\begin{equation}
  \Delta({\bf k}^{2})=\int_{m_{t}^{2}}^{\infty}
     \frac{\rho(m^{\prime2})dm^{\prime2}}{{\bf k}^{2}+m^{\prime2}} \ ,
\end{equation}
with mass distribution
\FL
\begin{equation}
  \rho(m^{\prime2})=\frac{1}{\pi}\frac{\gamma(m^{\prime2}-
      m_{t}^{2})^{n+1/2}} {(m^{\prime2}-m^{2})^{2}+\gamma^{2}
      \left(\frac{m^{\prime2}}{m^{2}}\right)^{2n}
      (m^{\prime2}-m_{t}^{2})^{2n+1}}  \ ,
\end{equation}
and where
\begin{equation}
  \gamma=m\Gamma\,(m^{2}-m_{t}^{2})^{-(n+1/2)} \ .
\end{equation}
Here $\Gamma$ denotes the width and $n=0,1$ for spin-0 and spin-1 
mesons, respectively~\cite{Sch71}. The charged $\rho$ meson decays
into a neutral and a charged pion and the threshold mass is
$m_{t}=m_{\pi^{0}}+m_{\pi^{\pm}}$. The neutral $\rho$ meson cannot 
decay into two neutral pions and it decays into two charged pions, 
and so now the threshold mass is $m_{t}=2m_{\pi^{\pm}}$. 
The $\epsilon$ meson is an isoscalar meson which decays into both
two neutral or two charged pions in the ratio 1:2. In our present 
models these distinctions have been explicitly accounted for, which
is another extension of the old Nijm78 model. 

The configuration-space potential due to the exchange of a broad meson 
is calculated exactly. This exact potential is then approximated by 
the sum of two potentials of stable mesons~\cite{Bin71}
\begin{eqnarray}
  \int_{m_{t}}^{\infty}dm'2m'&&\rho(m^{\prime2})
    m^{\prime} \phi^{0}_{C}(m',r)                          \nonumber\\
    \approx && \beta_{1}m_{1}\phi^{0}_{C}(m_{1},r) +
               \beta_{2}m_{2}\phi^{0}_{C}(m_{2},r) \ . \label{Yukfit}
\end{eqnarray}
Fitting from 0--2 fm yields the values as given in Table~\ref{broad}.

\subsection{Regularized Reid potential}
\label{subsec:Reid}
A disadvantage of the original Reid68 potential is that, at the time
of its construction, the quality of the $np$ data was very poor. As a
consequence, the Reid68 potential can no longer properly describe
the numerous new and much more accurate $np$ data. Another disadvantage
is that the Reid68 potential has an $r^{-1}$ singularity in all
partial waves. Here we present an updated version of the Reid potential, 
where these singularities have been removed via the inclusion of a 
dipole form factor. With this choice, the tensor potential now also 
vanishes at the origin, as it should. 

As is the case for the original Reid68 potential, the OPE potential is 
explicitly included, while we now account for the neutral-pion and 
charged-pion mass differences as in Eqs.~(\ref{VOPE})--(\ref{Vnp}).
For the pion-nucleon coupling constant at the pion pole we 
take~\cite{Klo91,St93c} $f_{\pi}^{2}=0.075$, and for the dipole cutoff 
parameter we choose $\Lambda=8m_{\pi}$.
In the OPE potential (\ref{VOPE}) we use $\phi^{1}_{C}$ only for the
$S$ waves. For all other partial waves, we found it more convenient
to use $\phi_{C}^{0}$ instead of $\phi_{C}^{1}$. 
Note that $\phi_{C}^{1}$ equals $\phi_{C}^{0}$ up to a modified 
$\delta$ function [see Eq.~(\ref{phi1C})], and that this modified 
$\delta$ function is screened by the centrifugal barrier for all 
these other partial waves, except the $S$ waves.

Starting with this OPE potential, the potential in each partial wave 
can now be extended by choosing a convenient combination of
central, tensor, and spin-orbit functions with arbitrary masses and
cutoff parameters. In the construction presented in the following we 
(more or less arbitrarily) settled for integer multiples of a mean
pion mass $\overline{m}=(m_{\pi^{0}}+2m_{\pi^{\pm}})/3$, while the 
cutoff mass in the dipole form factor is chosen to be 
$\Lambda=8\overline{m}$ everywhere. 
For notational reasons we next define 
\[   \begin{array}{l}
        Y(p)=p\overline{m}\,\phi^{0}_{C}(p\overline{m},r) , \\[0.2cm]
        Z(p)=p\overline{m}\,\phi^{0}_{T}(p\overline{m},r) , \\[0.2cm]
        W(p)=p\overline{m}\,\phi^{0}_{SO}(p\overline{m},r) ,
     \end{array}
\]
with $p$ an integer and $\phi^{0}_{X}$ given by Eqs.~(\ref{phidipole})
and (\ref{phiTSO}).
For the coefficients multiplying these functions, we use $A_{ip}$ for 
the isovector potentials, whereas the coefficients $B_{ip}$ are for 
the isoscalar and $np$ $^{1}S_{0}$ potentials. The index $i$ subsequently
labels the different partial waves. For the total potential in a 
particular partial wave one should, of course, add the appropriate 
OPE potential as given by Eqs.~(\ref{VOPE})--(\ref{Vnp}).
 
For the non-OPE parts in the isovector singlet partial waves
($I=1$, $S=0$, $L=J$) we use
\begin{eqnarray}
  V_{pp}(^{1}S_{0}) &=& A_{12}Y(2)+A_{13}Y(3)+A_{14}Y(4)    \nonumber\\
                    & &+A_{15}Y(5)+A_{16}Y(6)            \ , \nonumber\\
  V_{np}(^{1}S_{0}) &=& B_{13}Y(3)+B_{14}Y(4)+B_{15}Y(5)
                       +B_{16}Y(6)                       \ , \nonumber\\
  V(^{1}D_{2}) &=& A_{24}Y(4)+A_{25}Y(5)+A_{26}Y(6)   \ , \label{sin1}\\
  V(^{1}G_{4}) &=& A_{33}Y(3)                            \ , \nonumber\\
  V(^{1}J_{J}) &=& V_{pp}(^{1}S_{0})\ \ \ {\rm for}\ J\geq6 \ , \nonumber
\end{eqnarray}
where the distinction between the $pp$ and $np$ ${^{1}S}_{0}$
potentials is necessary because of the well-known breaking of charge
independence in the $pp$ and $np$ ${^{1}S}_{0}$ partial waves.
The coefficients $A_{ip}$ and $B_{ip}$ are to be fitted.
The presence of the two-pion range piece $A_{02}Y(2)$ in the $pp$
${^{1}S}_{0}$ potential is purely coincidental, and was only included to
improve the quality of the fit. A similar term in the $np$ ${^{1}S}_{0}$
was much less effective, and so we decided to leave it out.

For the non-OPE parts in the isoscalar singlet partial waves
($I=0$, $S=0$, $L=J$) we use
\begin{eqnarray}
  V(^{1}P_{1}) &=& B_{23}Y(3)+B_{24}Y(4)+B_{25}Y(5)+B_{26}Y(6) \ , \nonumber\\
  V(^{1}F_{3}) &=& B_{33}Y(3)+B_{35}Y(5)               \ , \label{sin0}\\
  V(^{1}J_{J}) &=& V(^{1}P_{1})\ \ \ {\rm for}\ J\geq5 \ . \nonumber
\end{eqnarray}
For the isovector triplet uncoupled partial waves
($I=1$, $S=1$, $L=J$) we use
\begin{eqnarray}
  V(^{3}P_{0}) &=& A_{43}Y(3)+A_{45}Y(5)+A_{4z3}Z(3)    \ , \nonumber\\
  V(^{3}P_{1}) &=& A_{53}Y(3)+A_{55}Y(5)+A_{5z3}Z(3)    \ , \label{trip1}\\
  V(^{3}F_{3}) &=& A_{63}Y(3)                           \ , \nonumber
\end{eqnarray}
and the isoscalar triplet uncoupled partial waves
($I=0$, $S=1$, $L=J$) are parametrized as
\begin{eqnarray}
   V(^{3}D_{2}) &=& B_{43}Y(3)+B_{45}Y(5)+B_{4z3}Z(3)   \ ,  \nonumber\\
   V(^{3}G_{4}) &=& B_{53}Y(3)                       \ .   \label{trip0}
\end{eqnarray}
Following the parametrization of the original Reid68 potential, the
non-OPE potential in the triplet coupled partial waves
($S=1$, $L=J\pm1$) is parametrized as
\begin{equation}
    V=V_{C}+V_{T}S_{12}+V_{SO}{\bf L\cdot S}   \ , \label{copgen}
\end{equation}
where the isovector ($I=1$) potentials are given by
\begin{eqnarray}
  V_{C} &=& A_{73}Y(3)+A_{74}Y(4)+A_{75}Y(5)+A_{76}Y(6)  \ , \nonumber\\
  V_{T} &=& A_{7z4}Z(4)+A_{7z6}Z(6)                      \ , \nonumber\\
  V_{SO}&=& A_{7w3}W(3)+A_{7w5}W(5)\ \ \ {\rm for}\ J=2  \ , \label{cop1}\\
  V_{SO}&=& A_{8w3}W(3)           \ \ \ {\rm for}\ J=4   \ , \nonumber
\end{eqnarray}
and the isoscalar ($I=0$) potentials read
\begin{eqnarray}
  V_{C} &= & B_{62}Y(2)+B_{63}Y(3)+B_{64}Y(4)                 \nonumber\\
        &  &+B_{65}Y(5)+B_{66}Y(6)                       \ ,  \nonumber\\
  V_{T} &= & B_{6z4}Z(4)+B_{6z6}Z(6)                  \ ,  \label{cop0}\\
  V_{SO}&= & B_{6w3}W(3)+B_{6w5}W(5)\ \ \ {\rm for}\ J=1 \ ,  \nonumber\\
  V_{SO}&= & B_{7w3}W(3)+B_{7w5}W(5)\ \ \ {\rm for}\ J=3 \ .  \nonumber
\end{eqnarray}

Finally, for the triplet isovector partial waves ($I=1$, $S=1$) 
with $J\geq5$ we use Eq.~(\ref{copgen}) with the central and tensor 
potentials of Eq.~(\ref{cop1}), and the spin-orbit potential equal 
to zero. Similarly, for the triplet isoscalar partial waves
($I=0$, $S=1$) with $J\geq5$ we use the central and tensor potentials 
of Eq.~(\ref{cop0}).
This choice is analogous to the extension of the Reid68 potential
to the higher partial waves as given by Day~\cite{Day81}.

\section{RESULTS}
\label{results}
The parameters of the potential models are optimized by minimizing
the $\chi^{2}$ in a direct fit to the data. Since the
scattering data are spread over a large number of energies (about
200 different energies for the $pp$ data and almost 400 different 
energies for the $np$ data in the 0--350 MeV energy range) and the
phase parameters need to be calculated up to at least $J\approx6$, the 
Schr\"odinger equation then has to be solved a very large number of 
times. This approach, therefore, is not very practical when the model 
parameters do not yet have reasonable values and, consequently, the 
$\chi^{2}$ is still very high. A more convenient approach is to start 
with the Nijmegen representation~\cite{St93b} of the $\chi^{2}$ 
hypersurface of the scattering data. It is obtained from the 10 
single-energy analyses and consists of 10 sets of phase parameters 
and the error matrix, each at a different energy.
The error matrix is the inverse of half the second-derivative matrix 
of the $\chi^{2}$ hypersurface with respect to the phase parameters 
up to $J=4$ within the energy bin of the single-energy analysis. 
This $\chi^{2}$ hypersurface is, in principle, independent of the 
particular partial-wave analysis. In practice the representation we
use is somewhat dependent on the Nijmegen multienergy analysis. The 
crucial point is, however, that it provides a very good and concise 
representation of the scattering data. For each change in the model 
parameters we need to solve the Schr\"odinger equation for all partial 
waves up to $J=4$ at only 10 different energies, which allows for a 
much quicker optimization for the parameters of the potential model.

In the last stage of the fitting procedure the potential parameters 
have been further optimized in a very time-consuming direct fit to 
the data. In this case we use the potential model in all partial waves 
with $J\leq6$, whereas in the higher partial waves we include OPE only. 
The final $\chi_{\rm min}^{2}$ of the potential should be obtained 
from this direct comparison with the experimental data.

\subsection{Nijm92pp}
Our first improvement of the Nijm78 potential~\cite{Nag78} was 
already started several years ago, when we constructed an update to 
the $pp$ data of the Nijm78 potential. This potential has been used 
in the Nijmegen analyses~\cite{Klo91,Klo92,St93c} to parametrize the 
$np$ isovector lower partial waves (except the $np$ $^{1}S_{0}$). 
In our latest analysis~\cite{St93b} (PWA93) we refer to it as the 
Nijm92$pp$ potential.
We found that a good fit to the $pp$ data could be obtained using
one cutoff parameter $\Lambda=827.53$ MeV for all three types of 
meson exchanges. Some of the coupling constants were not refitted, but 
were kept at the values of the original Nijm78 potential. The reason 
is that, when we only fit to the $pp$ data, we cannot incorporate the 
isospin dependence of the isovector exchanges (there are only $I=1$ 
partial waves). A direct comparison of this Nijm92$pp$ potential with 
the $pp$ data yields $\chi^{2}=2487.1$ for 1787 data, which means 
$\chi^{2}/N_{pp}=1.4$.

\subsection{Nijm93}
Now that the Nijmegen analysis of the $np$ data is also 
finished~\cite{St93b} and we have carefully scrutinized the $np$
database, we can extend the update of the Nijm78 potential to include 
the fit to the $np$ scattering data as well. 
This model we refer to as Nijm93.

As already mentioned in Sec.~\ref{subsec:Reid}, the $np$ $^{1}S_{0}$ 
partial wave has to be parametrized separately. The reason is that 
there is clear evidence for breaking of charge independence in the 
$^{1}S_{0}$ scattering lengths $a_{pp}$ and $a_{np}$. This difference 
in scattering lengths carries over into an approximately $2^{\circ}$ 
phase-shift difference between the $pp$ and $np$ $^{1}S_{0}$ phase shifts 
at higher energies. This difference cannot be explained as being only 
due to the difference between the $pp$ and $np$ OPE potentials. 
Allowing for a different value for the neutral-pion and charged-pion 
coupling constants does not help either, because the scattering lengths 
are very insensitive to these kind of changes.
To accommodate the $pp$ and $np$ $^{1}S_{0}$ differences, we therefore 
introduce a purely phenomenological breaking of charge independence 
between the $\rho^{0}$ and $\rho^{\pm}$ coupling constants. This breaking 
of charge independence is {\it only\/} assumed in the $^{1}S_{0}$ partial 
wave; for all other partial waves the $\rho^{0}$ and $\rho^{\pm}$
coupling constants are taken to be the same. 

The parameters for the Nijm93 potential, rounded to four or five 
significant figures, can be found in Table~\ref{parnym93}, where the
meson masses are the masses as listed by the 1990 Particle Data 
Group~\cite{PDG90}. The coupling constants are the values at
${\bf k}^{2}=0$. The pion coupling constants are fixed at $f^{2}=0.075$ 
at the corresponding pion pole ${\bf k}^{2}=-m^{2}_{\pi^{0}}$
or ${\bf k}^{2}=-m^{2}_{\pi^{\pm}}$; hence the different entries
for neutral- and charged-pion exchange at ${\bf k}^{2}=0$.
The $\epsilon$ and Pomeron coupling constants are rather large,
whereas the $\omega$ coupling constant is reasonably small.
For the $\rho$ coupling constants we find $(f/g)_{\rho}=4.094$, which 
is close to the value 3.7 from naive vector-meson dominance of the
isovector electromagnetic form factors of the nucleon.

For the Nijm93 potential we find $\chi^{2}(pp)=3175.6$ for 1787 $pp$ 
data and $\chi^{2}(np)=4848.4$ for 2514 $np$ data. So for the $pp$ 
data $\chi^{2}/N_{pp}=1.8$, for the $np$ data $\chi^{2}/N_{np}=1.9$, 
and for all $N\!N$ data $\chi^{2}/N_{\rm data}=1.87$. 
We find that this 15-parameter conventional meson-exchange potential 
cannot do better than $\chi^{2}/N_{\rm data}\approx1.87$, a result which
is also found for similar potential models such as the Paris80
and all the Bonn potentials. Apparently, the conventional meson-exchange
potentials cannot compete in quality with the Nijmegen PWA93. 
This indicates that these models lack some important physics.

\subsection{Nijm~I and Nijm~II}
In order to be able to construct a potential model which is of almost 
the same quality as the Nijmegen PWA93 ($\chi^{2}/N_{\rm data}\approx1$),
we follow a different approach and take advantage of the success of 
the Reid68 potential. We expect that, when we start with the 
Nijmegen potential (which already has a reasonable $\chi^{2}$ on 
the $pp$ data), we can construct a Reid-like potential where 
in each partial wave we probably need to adjust only a few parameters 
in order to arrive at $\chi^{2}/N_{\rm data}\approx1$.
The potential forms are then given by a set of slightly adjusted 
Nijmegen potentials, each representing one particular partial wave. 
Starting with the parameters of the Nijm92$pp$ potential~\cite{remark},
we find that for most partial waves an adjustment of the $f_{\rho}$ 
and $g_{\epsilon}$ coupling constants already gives very good results. 
Counting all parameters which have been adjusted in the fit of each 
partial wave, we arrive at a total of 41 parameters. This should be 
compared with the 39 parameters used in the Nijmegen PWA93.

In the last stage, the parameters of this Reid-like potential are 
optimized in a direct fit to the data. The potential is then used in
all partial waves up to $J=6$ simultaneously. This model we refer to 
as Nijm~I. It has $\chi^{2}(pp)=1795.8$ and $\chi^{2}(np)=2627.3$, and
so $\chi^{2}/N_{\rm data}=1.03$ on all $pp$ and $np$ scattering data.
 
We have also constructed a local Reid-like Nijmegen potential, where 
we leave out the explicit momentum-dependent terms which give rise to
non-local contributions to the configuration-space potential as
expressed in Eq.~(\ref{nonlocal}). We follow the same procedure as 
for the non-local Nijm~I potential. First, the parameters
are adjusted in a fit to the representation of the $\chi^{2}$
hypersurface, and then further optimized in a direct fit to the data. 
For this local potential, denoted by Nijm~II, we use a total of 47 
parameters and we find $\chi^{2}(pp)=1795.8$ and $\chi^{2}(np)=2625.7$, 
and so $\chi^{2}/N_{\rm data}=1.03$.
 
Although these potentials are purely phenomenological (except for
the correct OPE tail) and the coupling constants have no physical
meaning, these potentials are the first to give an excellent
description of the $N\!N$ scattering data. They have already been used
successfully in three- and many-body calculations~\cite{Fri93,Zhe93}.

\subsection{Reid93}
Finally, we have constructed an updated Reid potential based on the 
original Reid68 potential. This regularized version, denoted by 
Reid93 and discussed in Sec.~\ref{subsec:Reid}, gives an equally good 
description of the data as do the Nijm~I and Nijm~II potentials. 
The 50 phenomenological potential parameters $A_{ij}$ and $B_{ij}$
were fitted to the data, resulting in $\chi^{2}(pp)=1795.1$ and 
$\chi^{2}(np)=2624.6$, and so also for this potential 
$\chi^{2}/N_{\rm data}=1.03$. This Reid93 potential has been used in
a triton calculation as well~\cite{Fri93}.

\subsection{Comparison of the potentials}
The results of the non-local Nijm~I potential, the local Nijm~II
potential, and the local Reid93 potential are summarized in
Table~\ref{compare}. The results of the Nijmegen PWA93 are shown for 
comparison. Although $\chi^{2}/N_{\rm data}$ for these three potentials 
is already very good, their description of the $np$ data is still not as 
good as the description of these $np$ data in the partial-wave analysis. 
Here we should also mention that we did not do a thorough investigation
into the minimum number of parameters required to get these results 
(as we did for the partial-wave analysis). The reason is that in order 
to do this properly, all potential parameters have to be fitted 
simultaneously to all $N\!N$ data. But we found it more successful to do 
a large number of fits where in each separate run only a (completely 
arbitrary) subset of the potential parameters was optimized. Therefore, 
we cannot rule out the possibility that an equally good fit can be 
obtained with a few parameters less.

As already mentioned in the Introduction, these new potentials
(except Nijm93) are in a sense alternative partial-wave analyses. 
The differences between the phase parameters of the potentials and
the phase parameters of the Nijmegen PWA93 are shown in Tables
\ref{phspp} and \ref{phsnp}. These differences provide an indication
for the systematic error on the results of the Nijmegen PWA93.
For the $np$ phase parameters the differences at some energies
(especially with the Reid93 potential) are relatively large.
However, one has to bear in mind that $\chi^{2}(np)$ of the
potential models is still substantially higher than that of the
multienergy partial-wave analysis.
On the other hand, the variation in the mixing parameter $\epsilon_{1}$ 
is small. It has often been claimed that this mixing parameter is ill 
determined and a wide range of values from potential models seemed 
acceptable (see, e.g., Ref.~\cite{Mac87,Mac89}). 
However, in our publication of the Nijmegen PWA93 we already
argued that $\epsilon_{1}$ is in fact known very accurately.
This is confirmed in Fig.~\ref{epsilon}, where we note that the
results of the Nijm~I and Nijm~II models lie essentially within the
{\it statistical\/} uncertainty as obtained in the Nijmegen PWA93.
Above 150 MeV, the result of the Reid93 model rises too strongly but 
is still within 2.5 standard deviations of the Nijmegen PWA93.

The potentials between 0 and 2 fm for the singlet, triplet uncoupled, 
and triplet coupled $np$ partial waves are shown in Figs.~\ref{potsin}, 
\ref{potunc}, and \ref{potcop}, respectively.
For the non-local potential Nijm~I we plot $V/(1+2\varphi)$, which 
more or less represents the effective potential when non-local terms 
are present (see Refs.~\cite{Gre62,Nag78}).
For coupled channels, the potential is a $2\times2$ matrix and the
$\varepsilon_{1}$ and $\varepsilon_{2}$ plots in Fig.~\ref{potcop}
represent the off-diagonal elements of the potential.
The main differences between the potentials show up in the inner 
region, i.e., for $r<1$ fm. In general, the non-local Nijm~I potential
is much softer than the local Nijm~II and Reid93 potentials, while the
Nijm~II potential is again much softer than the Reid93 potential.
The reason for this softness of the Nijmegen potentials is the 
exponential form factor.

Finally, all potential models have been fitted to the deuteron binding 
energy $B=2.224\,575(9)$ MeV~\cite{Leu82} using relativistic kinematics, 
i.e., 
\[
      B=M_{p}+M_{n}-\sqrt{M^{2}_{p}-\kappa^{2}}-
         \sqrt{M^{2}_{n}-\kappa^{2}}  \ ,
\]
rather than $B=\kappa^{2}/2M_{r}$. We also constructed versions 
of the Nijm~I and Nijm~II potentials to accommodate the latter 
nonrelativistic form. In any case the value $B=2.224\,575$ MeV is
exactly reproduced to this accuracy. 
Some of the other deuteron parameters are listed in Table~\ref{deuter}. 
The different potential models all give very similar results. 
Because we consider the potentials Nijm~I, Nijm~II, and Reid93 as 
alternative partial wave analyses, the values of the deuteron 
parameter $\eta$, $A_{s}$, and $N^{2}$ given by these potentials are 
new experimental determinations of these quantities. For the 
$D/S$-state ratio $\eta$ we find $\eta=0.0252(1)$ in good agreement 
with the recent determination by Rodning and Knutson~\cite{Rod90} of 
$\eta=0.0256(4)$. For the asymptotic $S$-state normalization $A_{S}$ 
we obtain $A_{s}=0.8843(10)$ fm$^{-1/2}$, which is in agreement with 
the determination by Kermode {\it et al.}~\cite{Ker83} of 
$A_{S}=0.8883(44)$ fm$^{-1/2}$.
This results in $N^{2}=A^{2}_{s}(1+\eta^{2})=0.7825(20)$ fm$^{-1}$.
However, there have been other experimental determinations of these
quantities which are not always in agreement with the values quoted
above. For a more complete list of these experimental determinations 
and a discussion of the differences between them, we refer to 
Ref.~\cite{Sto88}, and references cited therein.
A direct comparison of our value $Q_{d}=0.271(1)$ fm$^{2}$ of the 
deuteron quadrupole moment with the experimental value 0.2859(3) fm$^{2}$ 
of \cite{Bis79} is only possible after all possible corrections have 
been accounted for, which is outside the scope of this paper. 
However, we would like to turn the argument around and suggest that
these corrections must obviously be about 0.015~fm$^{2}$. It is quite
interesting to see that for our best potentials the $D$-state
probability is $P_{d}=5.665(30) \%$.
The deuteron parameters as well as the results for the scattering 
lengths for both the potential models and the Nijmegen PWA93 will be 
discussed in more detail elsewhere~\cite{Ter93}.
These potentials were used~\cite{Fri93} in calculations of the triton
binding energy. It turned out that all these potentials underbind
the triton by roughly 800 keV, a result which can be expected from
their $P_{d}$ values.

\section{CONCLUSIONS}
\label{conclusions}
In this paper we presented an update Nijm93 of the old Nijm78 $N\!N$
potential. It contains the correct OPE tail and has 
$\chi^{2}/N_{\rm data}=1.87$. Although it cannot compete in quality 
with the Nijmegen partial-wave analysis (a feature apparently all 
conventional meson-exchange potentials suffer from), the description 
of the $np$ data of the new Nijm93 model is substantially better than 
that of the original Nijm78 potential, which was fitted to the old 
1969 Livermore database~\cite{Mac69}.
Here we would also like to point out that in the Nijm93 model we 
do not include two-meson-exchange contributions such as $\pi\pi$ and 
$\pi\rho$ exchange, and we still get a reasonable description of the 
$^{1}P_{1}$ and $^{3}D_{2}$ phase shifts. This in contrast to claims 
made in the literature~\cite{Mac87,Mac89} that this is impossible.
However, this result is obtained at the cost of having a rather large 
value for the pseudoscalar (pion) cutoff mass of $\Lambda_{P}=1177.11$ 
MeV.

We have also presented three new high-quality $N\!N$ potentials. 
The Nijm~I potential is a non-local Reid-like potential where each of 
the lower partial waves up to $J=4$ is parametrized separately. For the 
higher partial waves we use the parameters of the Nijm92$pp$ potential, 
which is an update to the $pp$ data of the original Nijm78 potential. 
The Nijm~II potential is a local Reid-like potential, and does not 
contain any explicit momentum-dependent terms. Both potentials fit the 
$N\!N$ scattering data with a nearly optimal 
$\chi^{2}_{\rm min}/N_{\rm data}=1.03$.
A regularized update of the Reid potential, denoted by Reid93, gives
the same excellent $\chi^{2}_{\rm min}/N_{\rm data}=1.03$.

Computer codes for these Nijmegen potentials Nijm~I, Nijm~II, and Nijm93, 
and for the regularized Reid93 potential, in configuration space 
as well as in momentum space, can be readily obtained via anonymous FTP 
from thef-nym.sci.kun.nl.

\acknowledgments
We would like to thank Dr.\ J.L.\ Friar and Dr.\ Th.A.\ Rijken for
many helpful discussions.
Part of this work was included in the research program of the Stichting
voor Fundamenteel Onderzoek der Materie (FOM) with financial support
from the Nederlandse Organisatie voor Wetenschappelijk Onderzoek (NWO).
One of us (V.G.J.S.) would also like to thank the Australian Research
Council for financial support.

\narrowtext
\begin{table}
\caption{Values for the parameters of Eq.~(\protect\ref{Yukfit}) of 
         the two-pole approximation for the broad $\epsilon$ meson
         and the broad neutral and charged $\rho$ mesons. 
         Masses $m$ and widths $\Gamma$ are in MeV.}
\begin{tabular}{llll}
              & \multicolumn{1}{c}{$\epsilon$}  
              & \multicolumn{1}{c}{$\rho^{0}$}    
              & \multicolumn{1}{c}{$\rho^{\pm}$}      \\ \tableline
  $n$         &  0         &  1         &  1       \\
  $m$         &  760.0     &  768.7     &  768.3   \\
  $\Gamma$    &  640.0     &  152.4     &  149.1   \\
  $\beta_{1}$ &  0.16900   &  0.26552   &  0.38755 \\
  $m_{1}$     &  487.818   &  645.377   &  674.152 \\
  $\beta_{2}$ &  0.61302   &  0.56075   &  0.45083 \\
  $m_{2}$     &  1021.14   &  878.367   &  929.974
\end{tabular}
\label{broad}
\end{table}
 
\narrowtext
\setdec 0.00000
\begin{table}
\caption{Masses and meson-nucleon coupling constants at 
         ${\bf k}^{2}=0$ for the Nijm93 potential.
         For the $np$ $^{1}S_{0}$ partial wave the coupling
         constants of the charged-rho meson are increased
         by 4.371\% (see text). $P$ in the last line denotes
         the Pomeron. Note that $(f/g)_{\rho}=4.094$.} 
\begin{tabular}{lccc}
               & $m$\ (MeV)            & $g^{2}$      & $f^{2}$      \\ 
\tableline
 $\pi^{\pm}$   &\dec 139.5675          &              &\dec 0.07395  \\
 $\pi^{0}$     &\dec 134.9739          &              &\dec 0.07402  \\
 $\eta$        &\dec 548.8             &              &\dec 0.01514  \\
 $\eta'$       &\dec 957.5             &              &\dec 0.01466  \\
 $\Lambda_{P}$ &\dec 1177.11           &              &              \\
               &                       &              &              \\
 $\rho^{\pm}$  & 768.3, $\Gamma=149.1$ &\dec  0.8481  &\dec14.217    \\    
 $\rho^{0}$    & 768.7, $\Gamma=152.4$ &\dec  0.8481  &\dec14.217    \\    
 $\omega$      &\dec 781.95            &\dec  9.1765  &\dec 0.3383   \\
 $\phi$        &\dec1019.412           &\dec  0.0985  &      0       \\
 $\Lambda_{V}$ &\dec 904.50            &              &              \\
               &                       &              &              \\
 $a_{0}^{\pm}, a_{0}^{0}$ &\dec 983.3  &\dec  1.9174  &              \\
 $\epsilon$    & 760.0, $\Gamma=640.0$ &\dec 28.196   &              \\
 $f_{0}$       &\dec 975.6             &\dec 12.142   &              \\
 $\Lambda_{S}$ &\dec 554.40            &              &              \\
               &                       &              &              \\
 $a_{2}$       &\dec 208.16            &\dec  0.0486  &              \\
 $P,f_{2},f'_{2}$ &\dec 208.16         &\dec 27.339   &
\end{tabular}
\label{parnym93}
\end{table}
 
\narrowtext
\begin{table}
\caption{$\chi^{2}$ for the new potential models in comparison with
         the Nijmegen multienergy analysis~\protect\cite{St93b}
         PWA93. We also show the number of parameters 
         ($N_{\rm par}$) for each model.} 
\begin{tabular}{lrrrrr}
              &  PWA93 & Nijm I & Nijm II & Reid93 & Nijm93 \\
\tableline
       $pp$   & 1787.0 & 1795.8 & 1795.8  & 1795.1 & 3175.6 \\	
       $np$   & 2489.2 & 2627.3 & 2625.7  & 2624.6 & 4848.4 \\	
      Total   & 4276.2 & 4423.1 & 4421.5  & 4419.7 & 8023.9 \\
$N_{\rm par}$ &   39   &   41   &   47    &   50   &   15   \\
$\chi^{2}/N_{\rm data}$ 
              &  0.99  &   1.03 &   1.03  &   1.03 &   1.87
\end{tabular}
\label{compare}
\end{table}

\mediumtext
\begin{table}
\squeezetable
\caption{$pp$ phase shifts in degrees. For each energy the rows give 
         the values from the Nijmegen multienergy partial-wave 
         analysis~\protect\cite{St93b}, the non-local Nijm~I potential, 
         the local Nijm~II potential, and the Reid93 potential, 
         respectively.}
\begin{tabular}{rddddddd}
 $T_{\rm lab}$ & $^{1}S_{0}$ & $^{1}D_{2}$ & $^{3}P_{0}$ & $^{3}P_{1}$ 
               & $^{3}P_{2}$ & $\varepsilon_{2}$ & $^{3}F_{2}$  \\
   \tableline
   1 &  32.77 &   0.00 &   0.13 & --0.08 &   0.01 & --0.00 &   0.00 \\
     &  32.79 &   0.00 &   0.13 & --0.08 &   0.01 & --0.00 &   0.00 \\
     &  32.80 &   0.00 &   0.13 & --0.08 &   0.01 & --0.00 &   0.00 \\
     &  32.79 &   0.00 &   0.13 & --0.08 &   0.01 & --0.00 &   0.00 \\
   5 &  54.85 &   0.04 &   1.58 & --0.90 &   0.21 & --0.05 &   0.00 \\
     &  54.88 &   0.04 &   1.58 & --0.90 &   0.22 & --0.05 &   0.00 \\
     &  54.91 &   0.04 &   1.57 & --0.89 &   0.22 & --0.05 &   0.00 \\
     &  54.85 &   0.04 &   1.57 & --0.90 &   0.21 & --0.05 &   0.00 \\
  10 &  55.22 &   0.17 &   3.73 & --2.06 &   0.65 & --0.20 &   0.01 \\
     &  55.25 &   0.17 &   3.73 & --2.05 &   0.66 & --0.20 &   0.01 \\
     &  55.28 &   0.17 &   3.70 & --2.04 &   0.65 & --0.20 &   0.01 \\
     &  55.22 &   0.16 &   3.71 & --2.05 &   0.65 & --0.20 &   0.01 \\
  25 &  48.66 &   0.70 &   8.58 & --4.93 &   2.49 & --0.81 &   0.10 \\
     &  48.68 &   0.70 &   8.60 & --4.91 &   2.50 & --0.81 &   0.11 \\
     &  48.72 &   0.70 &   8.52 & --4.89 &   2.49 & --0.81 &   0.11 \\
     &  48.71 &   0.69 &   8.60 & --4.90 &   2.49 & --0.80 &   0.10 \\
  50 &  38.92 &   1.71 &  11.47 & --8.32 &   5.85 & --1.71 &   0.34 \\
     &  38.91 &   1.71 &  11.55 & --8.31 &   5.85 & --1.70 &   0.34 \\
     &  38.92 &   1.70 &  11.48 & --8.30 &   5.84 & --1.70 &   0.34 \\
     &  39.03 &   1.68 &  11.67 & --8.30 &   5.83 & --1.69 &   0.34 \\
 100 &  24.98 &   3.79 &   9.45 &--13.26 &  11.01 & --2.66 &   0.82 \\
     &  24.96 &   3.73 &   9.50 &--13.30 &  10.96 & --2.63 &   0.82 \\
     &  24.91 &   3.75 &   9.55 &--13.33 &  10.97 & --2.64 &   0.83 \\
     &  25.09 &   3.71 &   9.79 &--13.30 &  10.97 & --2.61 &   0.81 \\
 150 &  14.77 &   5.61 &   4.74 &--17.43 &  13.98 & --2.87 &   1.20 \\
     &  14.79 &   5.60 &   4.63 &--17.51 &  13.94 & --2.87 &   1.19 \\
     &  14.70 &   5.61 &   4.77 &--17.54 &  13.95 & --2.87 &   1.20 \\
     &  14.83 &   5.55 &   4.97 &--17.49 &  13.96 & --2.85 &   1.16 \\
 200 &   6.57 &   7.06 & --0.37 &--21.25 &  15.63 & --2.76 &   1.42 \\
     &   6.66 &   7.20 & --0.63 &--21.32 &  15.65 & --2.80 &   1.39 \\
     &   6.56 &   7.15 & --0.47 &--21.29 &  15.63 & --2.79 &   1.42 \\
     &   6.62 &   7.08 & --0.32 &--21.26 &  15.63 & --2.77 &   1.36 \\
 250 & --0.30 &   8.27 & --5.43 &--24.77 &  16.59 & --2.54 &   1.47 \\
     & --0.20 &   8.50 & --5.72 &--24.81 &  16.65 & --2.62 &   1.39 \\
     & --0.25 &   8.41 & --5.61 &--24.67 &  16.61 & --2.61 &   1.44 \\
     & --0.23 &   8.35 & --5.45 &--24.71 &  16.59 & --2.55 &   1.39 \\
 300 & --6.14 &   9.42 &--10.39 &--27.99 &  17.17 & --2.34 &   1.34 \\
     & --6.18 &   9.52 &--10.49 &--28.02 &  17.26 & --2.41 &   1.20 \\
     & --6.12 &   9.42 &--10.49 &--27.71 &  17.21 & --2.41 &   1.29 \\
     & --6.10 &   9.40 &--10.29 &--27.86 &  17.15 & --2.26 &   1.25 \\
 350 &--11.11 &  10.69 &--15.30 &--30.89 &  17.54 & --2.21 &   1.04 \\
     &--11.51 &  10.28 &--14.94 &--30.98 &  17.63 & --2.23 &   0.86 \\
     &--11.28 &  10.24 &--15.08 &--30.45 &  17.61 & --2.23 &   0.98 \\
     &--11.22 &  10.30 &--14.80 &--30.77 &  17.49 & --1.95 &   0.95
\end{tabular}
\label{phspp}
\end{table}

\widetext
\begin{table}
\squeezetable
\caption{$np$ phase shifts in degrees. For each energy the rows give 
         the values from the Nijmegen multienergy partial-wave 
         analysis~\protect\cite{St93b}, the non-local Nijm~I potential, 
         the local Nijm~II potential, and the Reid93 potential, 
         respectively.}
\begin{tabular}{rdddddddddddd}
 $T_{\rm lab}$ & $^{1}S_{0}$ & $^{3}P_{0}$ & $^{1}P_{1}$ & $^{3}P_{1}$ 
     & $^{3}S_{1}$ & $\varepsilon_{1}$ & $^{3}D_{1}$ & $^{1}D_{2}$ 
     & $^{3}D_{2}$ & $^{3}P_{2}$ & $\varepsilon_{2}$ & $^{3}F_{2}$ \\ 
\tableline
   1 &  62.07 &   0.18 & --0.19 & --0.11 & 147.75 &   0.11 & --0.01
                       &   0.00 &   0.01 &   0.02 & --0.00 &   0.00 \\
     &  62.11 &   0.18 & --0.19 & --0.11 & 147.76 &   0.10 & --0.01
                       &   0.00 &   0.01 &   0.02 & --0.00 &   0.00 \\
     &  62.09 &   0.18 & --0.19 & --0.11 & 147.75 &   0.10 & --0.01
                       &   0.00 &   0.01 &   0.02 & --0.00 &   0.00 \\
     &  61.89 &   0.18 & --0.19 & --0.11 & 147.73 &   0.10 & --0.01
                       &   0.00 &   0.01 &   0.02 & --0.00 &   0.00 \\
   5 &  63.63 &   1.63 & --1.49 & --0.94 & 118.18 &   0.67 & --0.18
                       &   0.04 &   0.22 &   0.25 & --0.05 &   0.00 \\
     &  63.74 &   1.62 & --1.50 & --0.93 & 118.19 &   0.67 & --0.18
                       &   0.04 &   0.22 &   0.25 & --0.05 &   0.00 \\
     &  63.66 &   1.60 & --1.52 & --0.93 & 118.17 &   0.66 & --0.18
                       &   0.04 &   0.22 &   0.25 & --0.05 &   0.00 \\
     &  63.23 &   1.61 & --1.48 & --0.93 & 118.15 &   0.66 & --0.18
                       &   0.04 &   0.22 &   0.26 & --0.05 &   0.00 \\
  10 &  59.95 &   3.65 & --3.04 & --2.06 & 102.61 &   1.16 & --0.68
                       &   0.16 &   0.85 &   0.71 & --0.18 &   0.01 \\
     &  60.10 &   3.64 & --3.08 & --2.05 & 102.62 &   1.15 & --0.68
                       &   0.16 &   0.85 &   0.71 & --0.18 &   0.01 \\
     &  59.99 &   3.61 & --3.11 & --2.04 & 102.59 &   1.13 & --0.67
                       &   0.16 &   0.85 &   0.71 & --0.18 &   0.01 \\
     &  59.46 &   3.64 & --3.04 & --2.05 & 102.59 &   1.14 & --0.67
                       &   0.16 &   0.85 &   0.72 & --0.18 &   0.01 \\
  25 &  50.90 &   8.13 & --6.31 & --4.88 &  80.63 &   1.79 & --2.80
                       &   0.68 &   3.71 &   2.56 & --0.76 &   0.09 \\
     &  51.04 &   8.16 & --6.42 & --4.86 &  80.59 &   1.77 & --2.80
                       &   0.69 &   3.72 &   2.57 & --0.75 &   0.09 \\
     &  50.88 &   8.09 & --6.51 & --4.84 &  80.56 &   1.73 & --2.80
                       &   0.68 &   3.72 &   2.56 & --0.75 &   0.09 \\
     &  50.41 &   8.24 & --6.37 & --4.85 &  80.63 &   1.74 & --2.75
                       &   0.67 &   3.73 &   2.61 & --0.75 &   0.09 \\
  50 &  40.54 &  10.70 & --9.67 & --8.25 &  62.77 &   2.11 & --6.43
                       &   1.73 &   8.97 &   5.89 & --1.63 &   0.30 \\
     &  40.56 &  10.81 & --9.80 & --8.25 &  62.64 &   2.09 & --6.45
                       &   1.72 &   8.98 &   5.88 & --1.61 &   0.31 \\
     &  40.35 &  10.76 & --9.96 & --8.24 &  62.62 &   2.00 & --6.45
                       &   1.72 &   8.97 &   5.87 & --1.62 &   0.31 \\
     &  40.18 &  11.13 & --9.89 & --8.26 &  62.78 &   2.03 & --6.31
                       &   1.69 &   9.00 &   6.00 & --1.60 &   0.30 \\
 100 &  26.78 &   8.46 &--14.52 &--13.24 &  43.23 &   2.42 &--12.24
                       &   3.90 &  17.27 &  10.94 & --2.58 &   0.76 \\
     &  26.44 &   8.57 &--14.42 &--13.30 &  42.98 &   2.44 &--12.26
                       &   3.83 &  17.26 &  10.89 & --2.54 &   0.76 \\
     &  26.18 &   8.65 &--14.59 &--13.33 &  42.95 &   2.25 &--12.31
                       &   3.85 &  17.22 &  10.88 & --2.54 &   0.77 \\
     &  26.32 &   9.22 &--14.91 &--13.38 &  43.18 &   2.36 &--12.07
                       &   3.79 &  17.12 &  11.21 & --2.54 &   0.74 \\
 150 &  16.93 &   3.69 &--18.65 &--17.46 &  30.72 &   2.75 &--16.49
                       &   5.79 &  22.12 &  13.84 & --2.80 &   1.12 \\
     &  16.33 &   3.67 &--18.23 &--17.56 &  30.47 &   2.83 &--16.45
                       &   5.77 &  22.15 &  13.80 & --2.79 &   1.11 \\
     &  16.02 &   3.83 &--18.32 &--17.59 &  30.40 &   2.59 &--16.61
                       &   5.78 &  22.05 &  13.78 & --2.78 &   1.12 \\
     &  16.11 &   4.47 &--18.91 &--17.68 &  30.67 &   2.82 &--16.30
                       &   5.68 &  21.72 &  14.27 & --2.82 &   1.08 \\
 200 &   8.94 & --1.44 &--22.18 &--21.30 &  21.22 &   3.13 &--19.71
                       &   7.29 &  24.50 &  15.46 & --2.70 &   1.33 \\
     &   8.27 & --1.60 &--21.51 &--21.41 &  21.08 &   3.27 &--19.62
                       &   7.43 &  24.61 &  15.48 & --2.73 &   1.28 \\
     &   7.92 & --1.42 &--21.52 &--21.38 &  20.98 &   3.03 &--19.85
                       &   7.38 &  24.49 &  15.42 & --2.71 &   1.31 \\
     &   7.83 & --0.74 &--22.15 &--21.55 &  21.31 &   3.40 &--19.46
                       &   7.26 &  23.96 &  16.01 & --2.79 &   1.27 \\
 250 &   1.96 & --6.51 &--25.14 &--24.84 &  13.39 &   3.56 &--22.21
                       &   8.53 &  25.40 &  16.39 & --2.49 &   1.35 \\
     &   1.48 & --6.69 &--24.28 &--24.93 &  13.46 &   3.70 &--22.13
                       &   8.78 &  25.48 &  16.47 & --2.56 &   1.26 \\
     &   1.10 & --6.55 &--24.22 &--24.78 &  13.35 &   3.55 &--22.34
                       &   8.68 &  25.46 &  16.39 & --2.54 &   1.32 \\
     &   0.86 & --5.80 &--24.71 &--25.08 &  13.81 &   4.05 &--21.84
                       &   8.55 &  24.81 &  17.02 & --2.61 &   1.28 \\
 300 & --4.46 &--11.47 &--27.58 &--28.07 &   6.60 &   4.03 &--24.14
                       &   9.69 &  25.45 &  16.95 & --2.30 &   1.19 \\
     & --4.43 &--11.46 &--26.52 &--28.17 &   7.00 &   4.10 &--24.21
                       &   9.82 &  25.32 &  17.08 & --2.36 &   1.06 \\
     & --4.84 &--11.43 &--26.43 &--27.85 &   6.90 &   4.12 &--24.26
                       &   9.73 &  25.52 &  16.99 & --2.35 &   1.14 \\
     & --5.17 &--10.57 &--26.69 &--28.31 &   7.55 &   4.74 &--23.64
                       &   9.63 &  24.84 &  17.64 & --2.36 &   1.12 \\
 350 &--10.58 &--16.39 &--29.66 &--30.97 &   0.50 &   4.57 &--25.57
                       &  10.96 &  25.08 &  17.31 & --2.18 &   0.87 \\
     & --9.70 &--15.90 &--28.25 &--31.14 &   1.36 &   4.47 &--26.00
                       &  10.60 &  24.48 &  17.44 & --2.19 &   0.69 \\
     &--10.10 &--16.01 &--28.17 &--30.60 &   1.30 &   4.74 &--25.74
                       &  10.59 &  25.02 &  17.40 & --2.18 &   0.81 \\
     &--10.47 &--15.02 &--28.15 &--31.27 &   2.20 &   5.45 &--25.00
                       &  10.55 &  24.40 &  18.03 & --2.09 &   0.82
\end{tabular}
\label{phsnp}
\end{table}

\narrowtext
\begin{table}
\caption{Deuteron properties of the potential models: $D/S$-ratio 
         $\eta$, asymptotic $S$-state normalization $A_{S}$ in
         ${\rm fm}^{-1/2}$, wave function normalization $N^{2}$ in 
         ${\rm fm}^{-1}$, $D$-state probability $P_{d}$ in \%, 
         and quadrupole moment $Q_{d}$ in ${\rm fm}^{2}$.}
\begin{tabular}{cllll}
              & Nijm I & Nijm II & Reid93  & Nijm93  \\
\tableline
      $\eta$  & 0.0253 & 0.0252  & 0.0251  & 0.0252  \\	
      $A_{S}$ & 0.8841 & 0.8845  & 0.8853  & 0.8842  \\	
      $N^{2}$ & 0.7821 & 0.7828  & 0.7843  & 0.7823  \\
      $P_{d}$ & 5.664  & 5.635   & 5.699   & 5.755   \\
      $Q_{d}$ & 0.2719 & 0.2707  & 0.2703  & 0.2706 
\end{tabular}
\label{deuter}
\end{table}

\begin{figure}
\caption{The mixing parameter $\epsilon_{1}$ of the various potentials
         and of the Nijmegen PWA93. The shaded band denotes the
         statistical error on $\epsilon_{1}$ as obtained in the
         Nijmegen partial-wave analysis.}
\label{epsilon}
\end{figure}

\begin{figure}
\caption{The new Nijmegen potentials in the $np$ singlet
         partial waves up to $J=3$.
         Solid line: Nijm~I, dashed line: Nijm~II, dotted line: Reid93.}
\label{potsin}
\end{figure}

\begin{figure}
\caption{The new Nijmegen potentials in the $np$ triplet
         uncoupled partial waves up to $J=3$.
         Solid line: Nijm~I, dashed line: Nijm~II, dotted line: Reid93.}
\label{potunc}
\end{figure}

\begin{figure}
\caption{The new Nijmegen potentials in the $np$ triplet
         coupled partial waves with $J=1$ and $J=2$.
         Solid line: Nijm~I, dashed line: Nijm~II, dotted line: Reid93.}
\label{potcop}
\end{figure}


\begin{references}
\bibitem{remark} At the time we first started to follow this approach, 
         the analysis of the $np$ data was still in progress. This is 
         the reason that the model we started with is the Nijm92$pp$
         potential, rather than the (more obvious) Nijm93 potential.
\bibitem{misprint} There are several misprints in Ref.~\cite{Nag78}:
         in Eq.~(6) the second expression should read
         ${\cal V}_{2}^{(V)}=-{\textstyle\frac{2}{3}}{\bf k}^{2}
         {\cal V}_{3}^{(V)}$; the $m^{2}$ in front of $\phi^{1}_{C}(r)$
         in Eq.~(15) should be dropped; the $\nabla^{2}$ operator in 
         Eq.~(16) should operate on $S_{12}$ as well, as 
         $[(-\nabla^{2})^{n}\phi^{0}_{T}(r)S_{12}]$; similarly, Eq.~(19) 
         should have $[(-\nabla^{2})^{n}\phi^{0}_{SO}(r){\bf L\cdot S}]$; 
         and the coefficient in front of the $Q_{12}$ operator in 
         Eq.~(34) should be divided by 4.
\end{references}
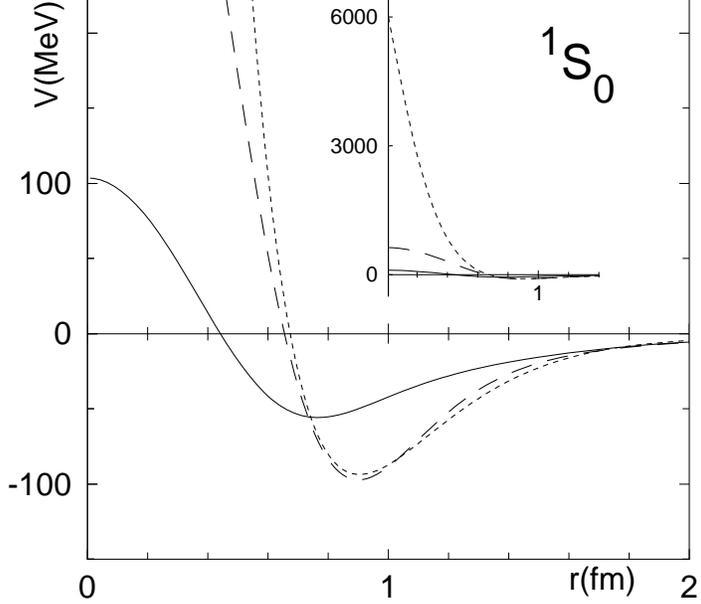
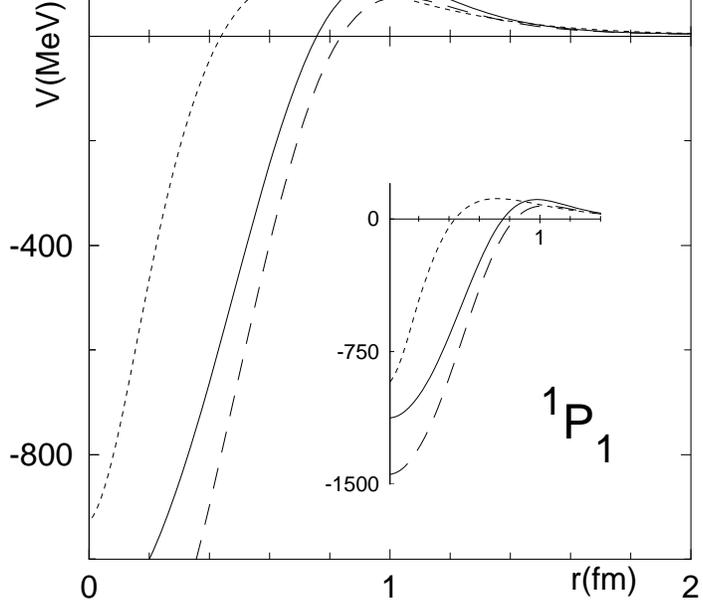
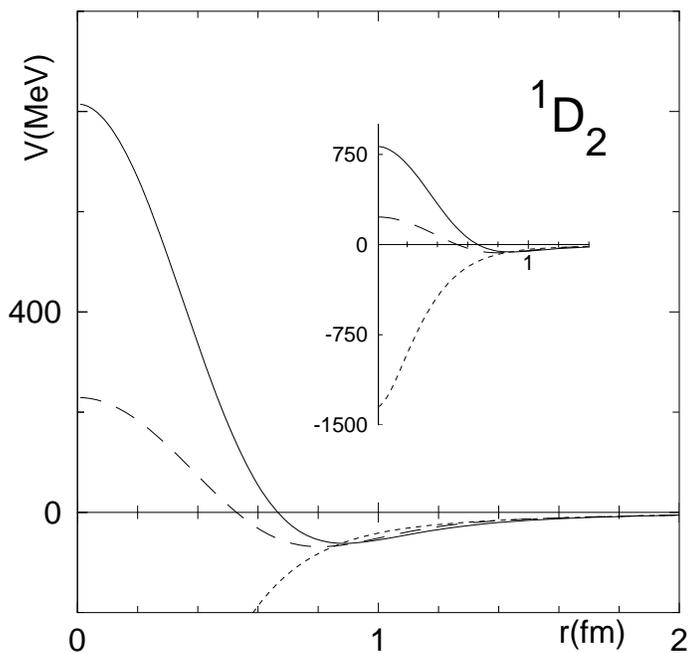
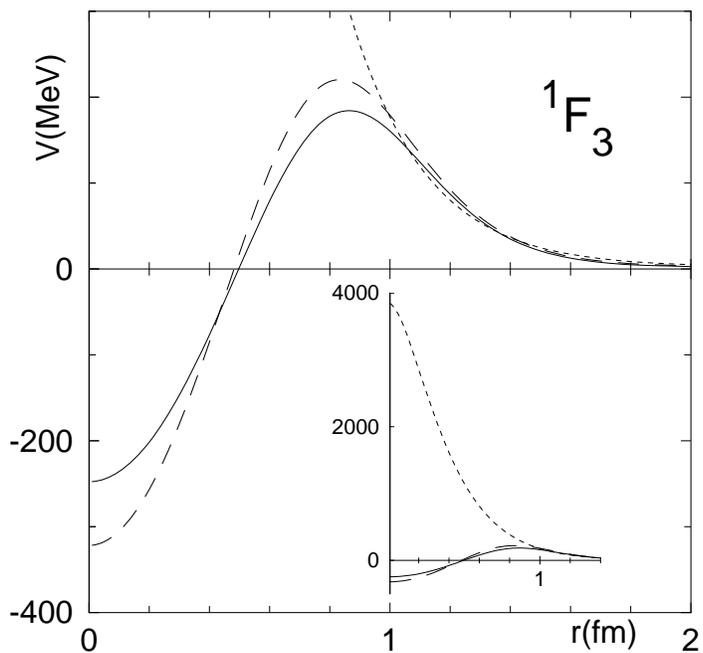
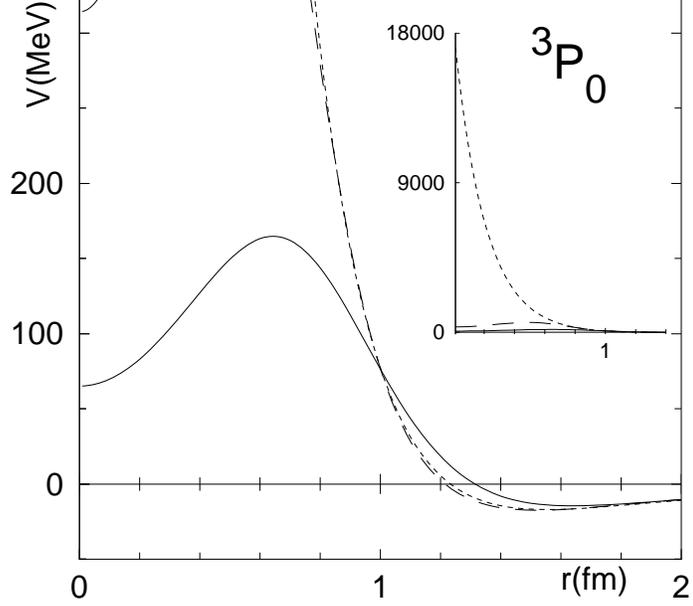
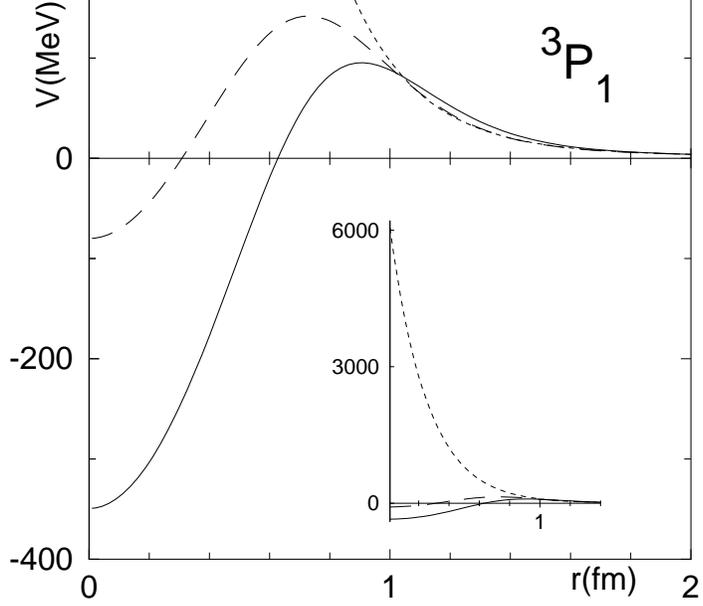
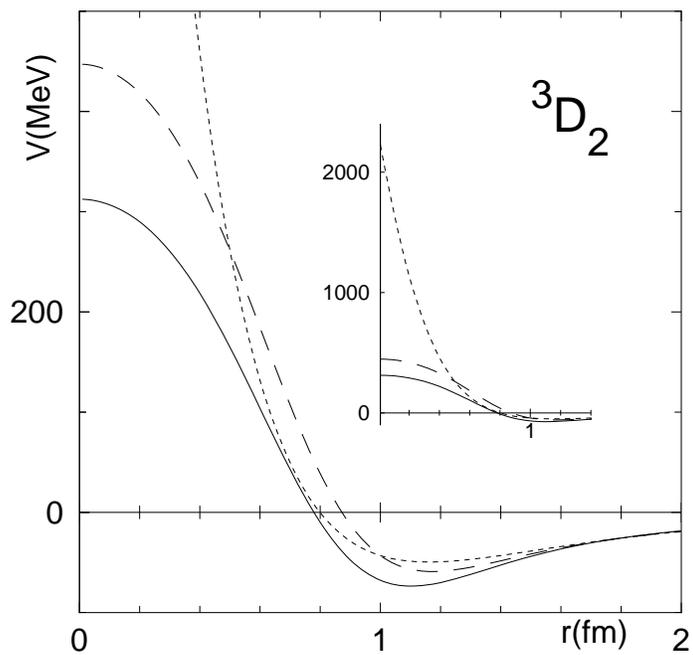
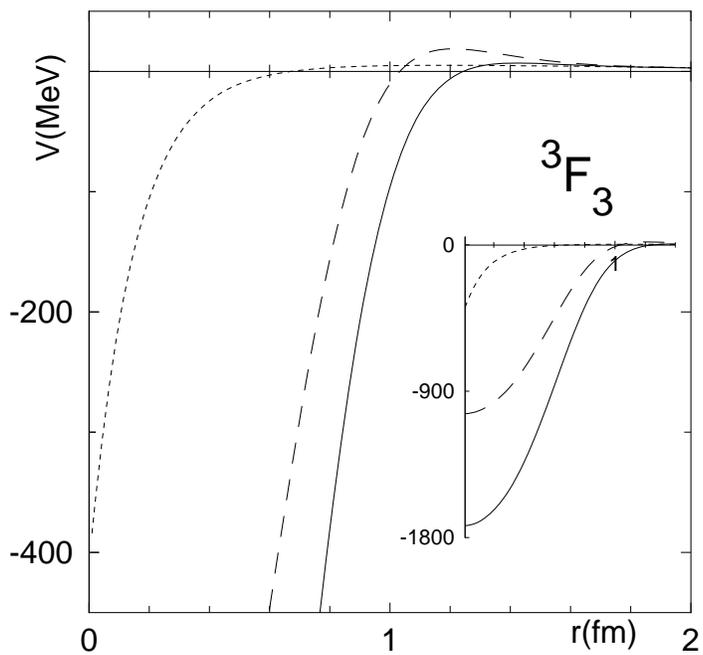
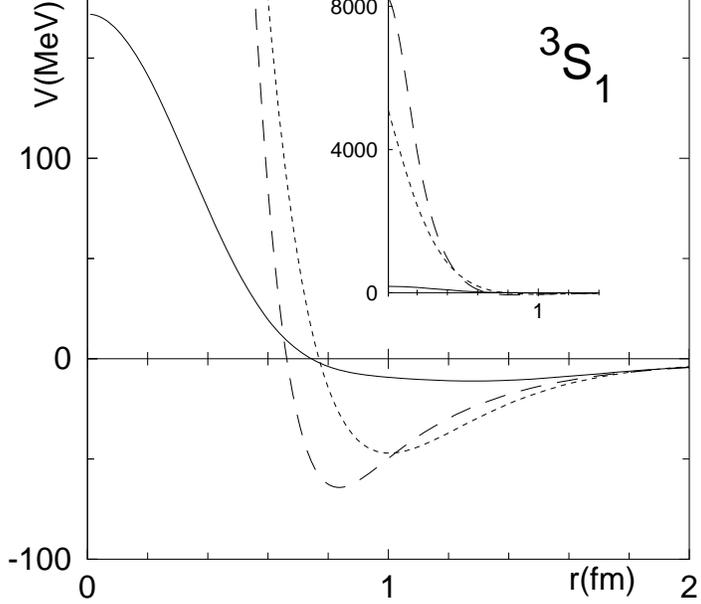
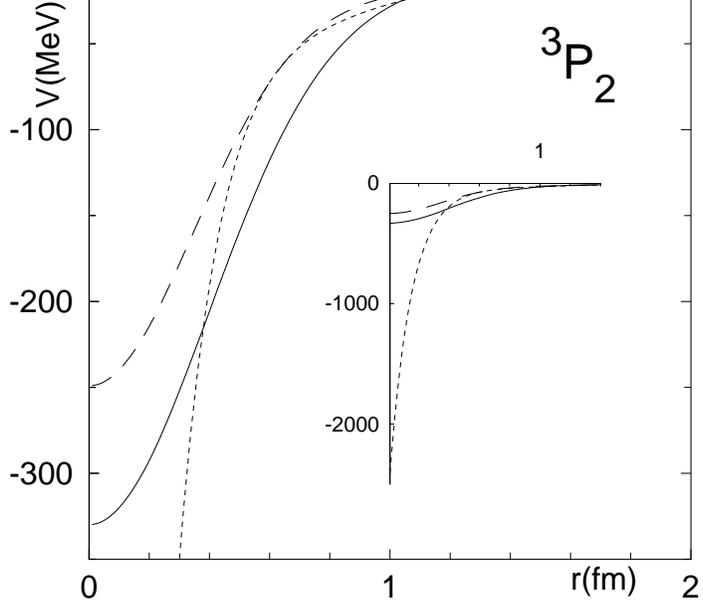
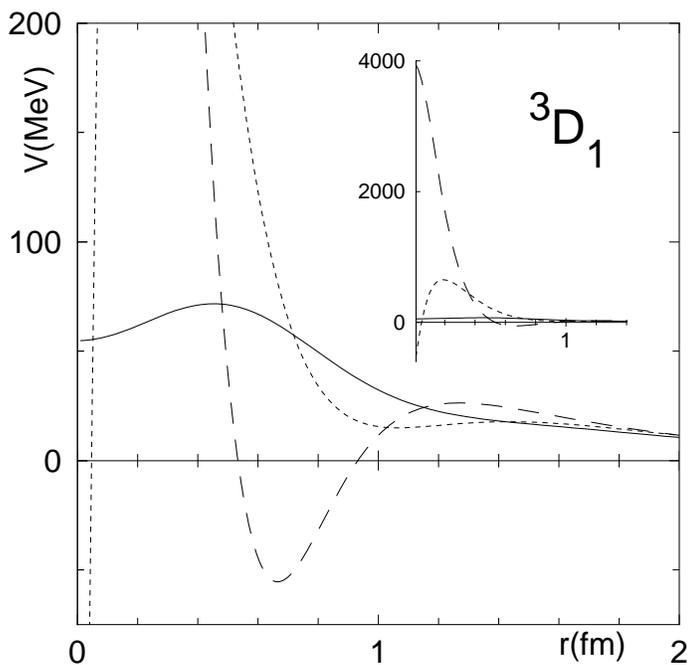
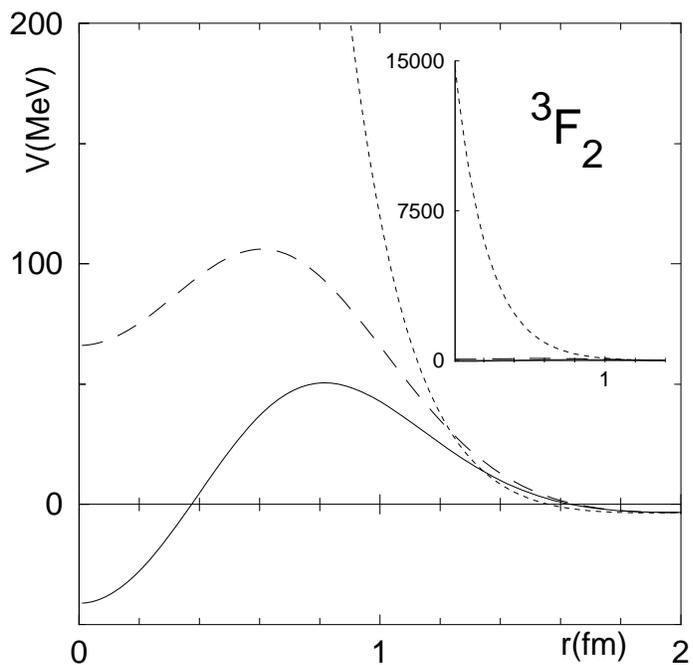
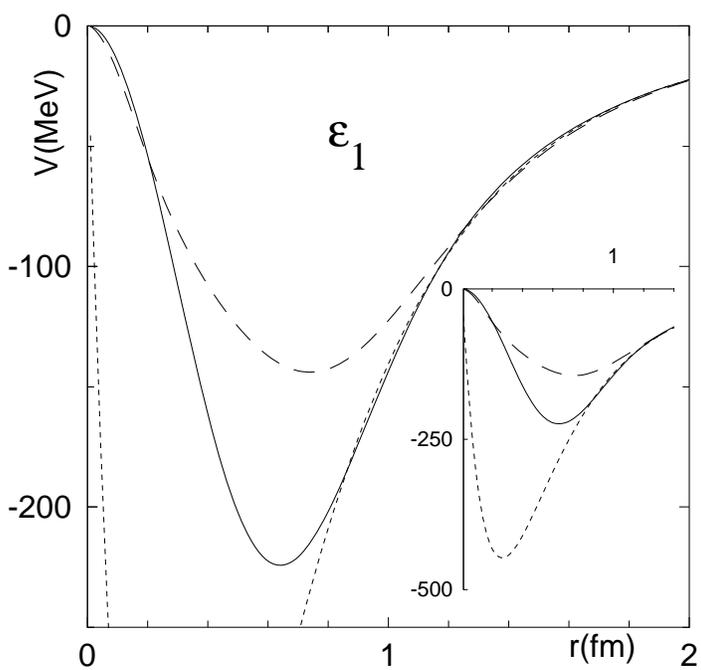
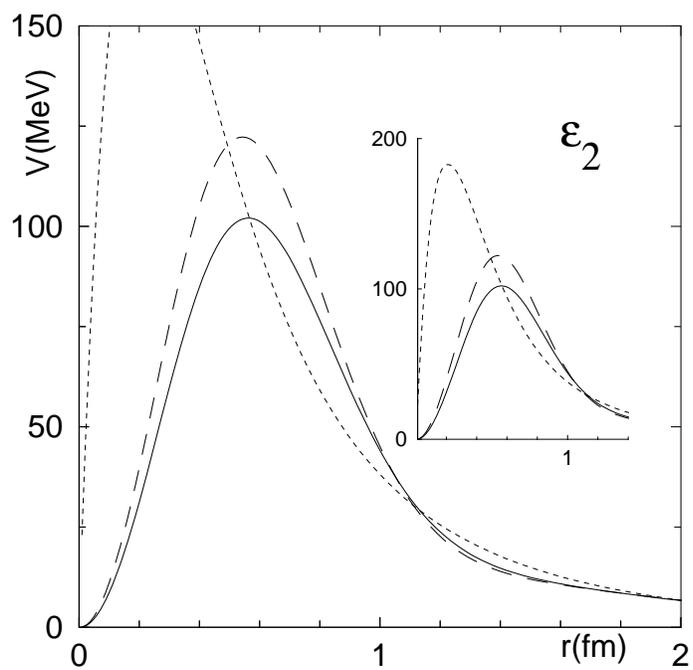
\end{document}